\documentclass[prd,aps,a4paper,nofootinbib,twocolumn]{revtex4-1} 

\newif\ifusesec
\usesectrue  
   
\usepackage{graphicx} 
\usepackage{mathrsfs}
\usepackage{latexsym,amsmath,amsfonts,amssymb}
\usepackage{multirow}

\newcommand{\beq}{\begin{equation}}
\newcommand{\eeq}{\end{equation}}

\begin{document}

\title{High post-Newtonian order gravitational self-force analytical results for \\ eccentric orbits around a Kerr black hole}

\author{Donato \surname{Bini}$^1$}
\author{Thibault \surname{Damour}$^2$}
\author{Andrea \surname{Geralico}$^1$}

\affiliation{$^1$Istituto per le Applicazioni del Calcolo ``M. Picone'', CNR, I-00185 Rome, Italy\\
$^2$Institut des Hautes Etudes Scientifiques, 91440 Bures-sur-Yvette, France}

\date{\today}

\begin{abstract}
We present the first analytic computation of the Detweiler-Barack-Sago gauge-invariant redshift function for a small mass in {\it eccentric} orbit around a {\it spinning} black hole. Our results give the redshift contributions that mix eccentricity and spin effects, through second order in eccentricity, 
second order in spin parameter, and the eight-and-a-half post-Newtonian order.
\end{abstract}

\pacs{04.20.Cv, 98.58.Fd}
\keywords{eccentric orbits, rotating black holes}
\maketitle

\section{Introduction}
 
The recent observation of the gravitational-wave signal emitted by a coalescing black-hole binary \cite{Abbott:2016blz}
reinforces the motivation for improving our theoretical description of the general relativistic dynamics of 
binary systems made of {\it spinning} bodies. Gravitational self-force computations of gauge-invariant
observables  \cite{Detweiler:2008ft,Barack:2009ey,Blanchet:2009sd,Damour:2009sm,Barack:2011ed} 
provide a mine of information which has recently shown its usefulness for informing the dynamical description of comparable-mass two-body problem \cite{Blanchet:2010zd,Barack:2010ny,LeTiec:2011ab,LeTiec:2011dp,Barausse:2011dq,Blanchet:2012at,Akcay:2012ea,Bini:2013zaa,Shah:2013uya,Bini:2013rfa,Dolan:2013roa,Bini:2014ica,Dolan:2014pja,Bini:2014zxa,Bini:2015mza,Bini:2015bla,Kavanagh:2015lva,Tiec:2015cxa,Bini:2015bfb,Hopper:2015icj,Akcay:2015pjz}.  
Up to now, the analytic dynamical information acquired from self-force computations has considered spin interactions and eccentric effects {\it separately},
without being able to mix them, i.e. without considering two-body interactions involving the product of powers of spin and of eccentricity.

Here, for the first time, we present an analytic computation of the self-force contribution $\delta U$ to the redshift function that include some cross-talk terms between spin and eccentricity.
We recall that the Detweiler-Barack-Sago \cite{Detweiler:2008ft,Barack:2011ed} (inverse) redshift function $U$ is defined as
\beq
\label{U}
U\left(m_2\Omega_r, m_2\Omega_\phi, \frac{m_1}{m_2}\right)= \frac{\displaystyle\oint dt}{\displaystyle\oint d\tau}=\frac{T_r}{{\mathcal T}_r} \, ,
\eeq
where all quantities refer to the perturbed spacetime metric (see Eq. \eqref{gperturbed} below). The (first-order) self-force contribution  
$\delta U$ is then defined as
\begin{eqnarray} \label{UdU}
&&U\left(m_2\Omega_r, m_2\Omega_\phi, \frac{a_2}{m_2},\frac{m_1}{m_2}\right)= U_0\left(m_2\Omega_r, m_2\Omega_\phi, \frac{a_2}{m_2}\right) \nonumber \\ 
&&+ \frac{m_1}{m_2} \delta U\left(m_2\Omega_r, m_2\Omega_\phi, \frac{a_2}{m_2}\right) + O\left( \frac{m_1^2}{m_2^2} \right)\,.
\end{eqnarray}
Here, $m_1$, $m_2$ (with $m_1 \leq m_2$, and, actually, $m_1 \ll m_2$ in our self-force computation) denote the masses of the binary system,
while  $a_2 \equiv S_2/m_2$ denotes the Kerr parameter of the larger mass. [The smaller mass $m_1$ is non spinning.] In addition,
$\Omega_r=2\pi/T_r$ and $\Omega_\phi=\Phi/T_r$ (where $T_r$ is the radial period and $\Phi$ the angular advance during one radial period)
denote the two fundamental frequencies of the orbital motion. The self-force contribution $\delta U$ is {\it a priori} defined as a function of the two
$m_2$-adimensionalized  fundamental frequencies of the orbit $m_2\Omega_r, m_2\Omega_\phi$ (and of the dimensionless spin parameter, $a_2/m_2$). It is, however, convenient to reexpress it as
a function of the eccentricity $e$ and dimensionless semi-latus rectum $p$ of the orbit, defined as
\begin{eqnarray} \label{ep}
e&=& \frac{r_{\rm apo}-r_{\rm peri}}{r_{\rm apo}+r_{\rm peri}}\,,\nonumber\\ 
p &=& \frac{2 \, r_{\rm peri}\, r_{\rm apo}}{m_2(r_{\rm peri}+r_{\rm apo})}\,,
\end{eqnarray}
where $p$ is dimensionless. We are interested here in (eccentric) bound orbits confined between a 
minimum radius ($r_{\rm peri}$) and a maximum one ($r_{\rm apo}$).
As usual, it is enough to know the link between $m_2\Omega_r, m_2\Omega_\phi$ 
and $e, p$ for the unperturbed motion, i.e. for an eccentric bound orbit in a Kerr background of mass $m_2$ and spin parameter $a_2$.
See, e.g., Ref. \cite{Glampedakis:2002ya} and references therein for a general discussion, and Sec. \ref{sec2} for explicit relations through second order in $e$ and the dimensionless spin parameter, that we shall henceforth denote as 
\beq
\hat a \equiv\frac{a_2}{m_2}\,.
\eeq
We shall work in the following with various terms in the expansion of $\delta U(p, e, \hat a)$
in powers of $e$ and $\hat a$: 
\begin{eqnarray} \label{dUea}
\delta U(u_p,e,\hat a) &=&\sum_{i,j=0}^\infty e^i  {\hat a}^j \delta U^{(e^i, a^j)}(u_p) \nonumber\\
&=& \delta U^{(e^0, a^0)}+e^2 \delta U^{(e^2, a^0)}+e^4 \delta U^{(e^4, a^0)}\nonumber\\
&+& \hat a \delta U^{(e^0, a^1)}+\hat a^2\delta U^{(e^0, a^2)}+\hat a^3\delta U^{(e^0, a^3)}\nonumber\\
&+& \hat a^4\delta U^{(e^0, a^4)}+\hat a^5\delta U^{(e^0, a^5)}+\hat a^6\delta U^{(e^0, a^6)}\nonumber\\
&+& e^2 \hat a\delta U^{(e^2, a^1)}+e^2 \hat a^2\delta U^{(e^2, a^2)}+\ldots\,,
\end{eqnarray}
where $u_p \equiv 1/p$.

The PN expansions (i.e., the expansions in powers of $u_p$) of the individual contributions $\delta U^{(e^i, a^j)}(u_p) $
that do not mix $e$ and $\hat a$ have been determined to high PN orders by recent analytic self-force computations. 
See Refs. \cite{Bini:2015bla,Kavanagh:2015lva} for $\delta U^{(e^0,a^0)}$;  Ref. \cite{Bini:2015bfb} for $\delta U^{(e^2,a^0)}$; Ref. \cite{Bini:2016qtx} for $\delta U^{(e^4,a^0)}$; and Refs. \cite{shah_capra2015,shah_MG14,Bini:2015xua,Kavanagh:2016idg} for $\delta U^{(e^0,a^j)}$, with $j \leq 6$. 
Note that higher order terms $\delta U^{(e^i,a^0)}$ in the eccentricity are known up to $i=20$, but at 4PN order only \cite{Akcay:2015pza,Hopper:2015icj,Bini:2016qtx}.
Let us only quote below, for illustration, some of the lowest-order PN coefficients, namely  
\begin{eqnarray}
-\delta U^{(e^0,a^0)}&=& u_p+2 u_p^2+5 u_p^3+\ldots \nonumber\\
-\delta U^{(e^2,a^0)}&=&-u_p-4 u_p^2-7 u_p^3+\ldots \nonumber\\
-\delta U^{(e^4,a^0)}&=&2u_p^2 -\frac14 u_p^3+\ldots \nonumber\\
-\delta U^{(e^6,a^0)}&=&\frac52  u_p^3+\ldots \nonumber\\
-\delta U^{(e^8,a^0)}&=&- \frac{15}{64}u_p^3+\ldots\nonumber\\
-\delta U^{(e^{10},a^0)}&=& -\frac{3}{64}u_p^3+\ldots\nonumber\\
-\delta U^{(e^{12},a^0)}&=&-\frac{5}{512} u_p^3+\ldots\nonumber\\
-\delta U^{(e^{14},a^0)}&=& -\left(-\frac{5}{12}+\frac{41}{4096}\pi^2\right) u_p^4+\ldots\nonumber\\
-\delta U^{(e^{16},a^0)}&=&\frac{45}{16384} u_p^3+\ldots\nonumber\\
-\delta U^{(e^{18},a^0)}&=&\frac{55}{16384}u_p^3+\ldots\nonumber\\
-\delta U^{(e^{20},a^0)}&=&\frac{429}{131072}u_p^3+\ldots\nonumber\\
-\delta U^{(e^0,a^1)}&=&-3 u_p^{5/2}-18 u_p^{7/2}-87 u_p^{9/2}+\ldots \nonumber\\
-\delta U^{(e^0,a^2)}&=&u_p^3+14 u_p^4+103 u_p^5+\ldots \nonumber\\
-\delta U^{(e^0,a^3)}&=&-3u_p^{9/2}-50u_p^{11/2}-445u_p^{13/2} +\ldots \nonumber\\
-\delta U^{(e^0,a^4)}&=&156u_p^7+8u_p^6 +\ldots \nonumber\\
-\delta U^{(e^0,a^5)}&=&-\left(\frac{512}{5}\zeta(3)-\frac{512}{5}\zeta(5)+\frac{46}{5}\right)u_p^{15/2} +\ldots \nonumber\\
-\delta U^{(e^0,a^6)}&=&-\left(- \frac{23072}{15}\zeta(3)+\frac{109184}{15}\zeta(5)+\frac{3292}{75}  \right. \nonumber\\
&&-\frac{28672}{5}\zeta(7)-\frac{856}{105}\pi^2+\frac{13696}{2625}\pi^4\nonumber\\
&& \left.  -\frac{219136}{496125}\pi^6  \right)u_p^9+\ldots  
\end{eqnarray}

In this work we shall analytically compute the PN expansions of the two eccentricity-spin-mixing contributions 
$\delta U^{(e^2,a^1)}(u_p)$ and $\delta U^{(e^2,a^2)}(u_p)$ through order $u_p^{9.5}$ (i.e. through 8.5PN order).

\section{Eccentric geodesic orbits in a Kerr spacetime}
\label{sec2}

Let us consider the (unperturbed) Kerr metric 
\begin{eqnarray}
\label{kerrmet}
ds^2&=&-\left(1-\frac{2Mr}{\Sigma}  \right) dt^2-\frac{4aMr \sin^2\theta}{\Sigma}dtd\phi+\frac{\Sigma}{\Delta}dr^2\nonumber\\
&+& \Sigma d\theta^2 +\left( r^2+a^2+\frac{2Mra^2\sin^2\theta}{\Sigma} \right)\sin^2\theta d\phi^2\,,
\end{eqnarray}
where   
\beq
\Delta= r^2+a^2-2Mr\,,\qquad 
\Sigma=r^2+a^2\cos^2\theta\,.
\eeq
For ease of notation, we sometimes denote  $m_2$ as $M$ and $a_2$ as $a$ (and, as above, $\hat a = a_2/m_2=a/M$).

Equatorial (timelike) geodesics are solutions of the equations 
\begin{eqnarray}
\label{geo_eq_1}
\frac{dt}{d\tau}&=&\frac{1}{r^2}\left[ -a(a\tilde E-\tilde L)+ \frac{r^2+a^2}{\Delta}P\right]\nonumber\\
\frac{dr}{d\tau}&=&\pm \frac{\sqrt{R}}{r^2} \nonumber\\
\frac{d\phi}{d\tau}&=&\frac{1}{r^2}\left[ \frac{a}{\Delta}P-( a\tilde E-\tilde L ) \right] \,,
\end{eqnarray}
with 
\begin{eqnarray}
P&=&\tilde E (r^2+a^2)-a\tilde L\,,\nonumber\\
R&=&P^2-\Delta [r^2+( a\tilde E-\tilde L )^2]\,,
\end{eqnarray}
where $\tau$ denotes the proper time parameter and $\tilde E$ and $\tilde L$ are the conserved energy and angular momentum per unit (reduced) mass.

As mentioned in the Introduction, we parametrize (unperturbed) bound orbits in terms of eccentricity $e$ and (dimensionless) semi-latus rectum $p$, Eqs. \eqref{ep}.
We will limit our considerations here to the second order approximation in both the dimensionless spin parameter $\hat a\equiv a/M$ and eccentricity $e$. In this case the functional links between $\tilde E$, or $\tilde L$, and ($p, e, \hat a$), are respectively given by
\begin{eqnarray}
\tilde E &=& \frac{1-2u_p}{(1-3u_p)^{1/2}}-\hat a \frac{u_p^{5/2}}{(1-3u_p)^{3/2}} +\hat a^2 \frac{ u_p^3}{2(1-3 u_p)^{5/2}} \nonumber\\
&& +e^2 \left[ \frac{(1-4u_p)^2 u_p}{2(1-3 u_p)^{3/2}(1-2u_p)} -\hat a \frac{(15 u_p-4) u_p^{5/2}}{2(1-3u_p)^{5/2}}   \right. \nonumber\\
&& \left. +\hat a^2 \frac{u_p^3(1-4u_p) (48 u_p^3-48 u_p^2+21 u_p-4)}{4(1-3 u_p)^{7/2}(1-2u_p)^2} \right]\nonumber\\
&&+O(\hat a^3, e^4)\,,
\end{eqnarray}
and
\begin{eqnarray}
\frac{\tilde L}{M} &=& \frac{1}{u_p^{1/2}(1-3u_p)^{1/2}}-\hat a \frac{3u_p(1-2u_p)}{(1-3u_p)^{3/2}} \nonumber\\
&+& \hat a^2 \frac{(6 u_p^2-5 u_p+2) u_p^{3/2}}{2(1-3 u_p)^{5/2}}  \nonumber\\
&+& e^2 \left[\frac{u_p^{1/2}}{2(1-3u_p)^{3/2}} -\hat a \frac{u_p(48u_p^3-16u_p^2-5u_p+2)}{2(1-3u_p)^{5/2}(1-2u_p)}  \right. \nonumber\\
&+& \left. \hat a^2 \frac{u_p^{3/2}(90u_p^3+15u_p^2-22u_p+4)}{4(1-3u_p)^{7/2}}  \right]+O(\hat a^3, e^4)\,,\nonumber\\
\end{eqnarray}
 where we recall that $u_p\equiv1/p$.

Up to order $e^2$ included,  the motion is explicitly given by 

\begin{widetext}

\begin{eqnarray}
\frac{r_0(t)}{ m_2}&=&
R_0 +e \, R_1(\cos \Omega_{r 0} t-1)+e^2 \, R_2(\cos(2\Omega_{r 0} t)-1)+O(\hat a^3,e^3)\,,\nonumber\\
\phi_0(t) &=&  \Omega_{\phi 0} t+ e \, \Phi_1\sin(\Omega_{r 0} t)+e^2 \, \Phi_2\sin(2\Omega_{r 0} t) +O(\hat a^3,e^3)\,,
\end{eqnarray}
where
\begin{eqnarray}
R_0 &=&  \frac{1+e+e^2}{u_p}\,,\nonumber\\
R_1 &=& \frac{1}{u_p}\,,\nonumber\\
R_2&=& -\frac{(1-11 u_p+26 u_p^2)}{2(1-6 u_p) (1-2 u_p) u_p} 
-\frac{u_p^{1/2}(1-u_p-32 u_p^2+108 u_p^3)}{ (1-2 u_p) (1-6 u_p)^2}\hat a\nonumber\\
&&
+\frac{u_p (-36 u_p^2+3 u_p+36 u_p^3+376 u_p^4-1728 u_p^5+2592 u_p^6+1)}{2 (1-6 u_p)^3 (1-2 u_p)^2}\hat a^2 \,,\nonumber\\
\Phi_1 &=& -\frac{2 (1-3 u_p)}{ (1-2 u_p) (1-6 u_p)^{1/2}}
+\frac{6 u_p^{3/2}(1-3 u_p+6 u_p^2)}{ (1-2 u_p) (1-6 u_p)^{3/2}}\hat a 
-\frac{(432 u_p^5-504 u_p^4+456 u_p^3-214 u_p^2+35 u_p+1) u_p^2}{ (1-2 u_p)^2 (1-6u_p)^{5/2}}\hat a^2\,,\nonumber\\
\Phi_2 &=&  -\frac{ (-250 u_p^2+300 u_p^3+64 u_p-5)}{4 (1-2 u_p)^2 (1-6 u_p)^{3/2}}
-\frac{u_p^{3/2}(511 u_p^2-1704 u_p^3-76 u_p+4+2196 u_p^4)}{2  (1-2 u_p)^2 (1-6 u_p)^{5/2} }\hat a \nonumber\\
&&
+\frac{(167616 u_p^7-214272 u_p^6+142704 u_p^5-68024 u_p^4+20856 u_p^3-3446 u_p^2+248 u_p-7) u_p^2}{8 (1-2 u_p)^3 (1-6 u_p)^{7/2}}\hat a^2\,.
\end{eqnarray}

The dimensionless orbital frequencies of the radial and azimuthal motions are respectively given by
\begin{eqnarray}
m_2 \Omega_{r0}&=&
(1-6u_p)^{1/2}u_p^{3/2}
+3u_p^3\frac{1+2u_p}{(1-6u_p)^{1/2}}\hat a
+\frac12u_p^{7/2}\frac{72u_p^3+24u_p^2-4u_p-3}{(1-6u_p)^{3/2}}\hat a^2\nonumber\\
&&
+e^2\left[
\frac34\frac{(266u_p^3-165u_p^2+32u_p-2)u_p^{3/2}}{(1-2u_p)(1-6u_p)^{3/2}}
+\frac34\frac{u_p^3(3180u_p^4-656u_p^3-403u_p^2+132u_p-10)}{(1-6u_p)^{5/2}(1-2u_p)}\hat a\right.\nonumber\\
&&\left.
-\frac38\frac{u_p^{7/2}(114336u_p^7-74688u_p^6-40u_p^5+6716u_p^4+272u_p^3-819u_p^2+168u_p-10)}{(1-6u_p)^{7/2}(1-2u_p)^2}\hat a^2
\right]
\nonumber\\
&&
+O(\hat a^3,e^3)\nonumber\\
m_2 \Omega_{\phi0}&=&u_p^{3/2}-u_p^3\hat a+u_p^{9/2}\hat a^2
+e^2\left[
-\frac32\frac{1+22u_p^2-10u_p}{(1-2u_p)(1-6u_p)}u_p^{3/2}
-\frac32\frac{1+4u_p-88u_p^2+264u_p^3}{(1-2u_p)(1-6u_p)^2}u_p^3\hat a\right.\nonumber\\
&&\left.
+\frac32\frac{1-7u_p+48u_p^2-330u_p^3+1564u_p^4-4296u_p^5+4752u_p^6}{(1-2u_p)^2(1-6u_p)^3}u_p^{7/2}\hat a^2
\right]
+O(\hat a^3,e^3)\,.
\end{eqnarray}

Finally, the (unperturbed) redshift variable $U_0=T_{r0}/{\mathcal T}_{r0}$ is given by
\begin{eqnarray} 
\label{U0}
U_0&=&
\frac1{(1-3u_p)^{1/2}}
-3\frac{u_p^{5/2}}{(1-3u_p)^{3/2}}\hat a
+\frac12\frac{u_p^3(1+6u_p)}{(1-3u_p)^{5/2}}\hat a^2\nonumber\\
&&
+e^2\left[
-\frac32\frac{(1+22u_p^2-10u_p)u_p}{(1-6u_p)(1-2u_p)(1-3u_p)^{3/2}}
+\frac32\frac{u_p^{5/2}(2-49u_p+346u_p^2-924u_p^3+792u_p^4}{(1-6u_p)^2(1-2u_p)(1-3u_p)^{5/2}}\hat a\right.\nonumber\\
&&\left.
+\frac34\frac{u_p^4(31-414u_p+2038u_p^2-4312u_p^3+3600u_p^4-2592u_p^5+4320u_p^6)}{(1-6u_p)^3(1-2u_p)^2(1-3u_p)^{7/2}}\hat a^2
\right]
+O(\hat a^3,e^3)\,.
\end{eqnarray}

\end{widetext}

\section{High PN-order analytical computation of the self-force correction to the averaged redshift function along eccentric orbits}

As already mentioned in the Introduction, we consider the first-order self-force correction to the 
Barack-Sago \cite{Barack:2011ed}  generalization to eccentric orbits of Detweiler's \cite{Detweiler:2008ft} circular, gauge-invariant first-order self-force correction to the (inverse) redshift. 
We denote this gauge-invariant measure of the $O(m_1/m_2)$ conservative self-force effect on eccentric orbits as $\delta U(m_2\Omega_r, m_2\Omega_\phi, a_2/m_2)= \delta U(p, e, \hat a)$, see Eq. \eqref{UdU}. 
It is given in terms of the $O(m_1/m_2)$ metric perturbation $h_{\mu\nu}$, where 
\beq \label{gperturbed}
g_{\mu\nu}(x^\alpha; m_1, m_2)=g^{(0)}_{\mu\nu}(x^\alpha; m_2)+\frac{m_1}{m_2} h_{\mu\nu}(x^\alpha)+O\left( \frac{m_1^2}{m_2^2} \right)
\eeq
[with $g^{(0)}_{\mu\nu}(x^\alpha; m_2, a_2)$ being the Kerr metric of mass $m_2$ and spin $m_2 a_2$] by the following
 time average
\beq
\label{delta_U1}
\delta U (u_p,e, \hat a)=\frac12 \, (U_{0})^2\langle h_{uk}\rangle_{t}\,.
\eeq
Here, we have expressed $\delta U$ (which is originally defined as a {\it proper} time $\tau$ average \cite{Barack:2011ed}) in terms of the {\it coordinate} time $t$ average of the mixed contraction $h_{uk}=h_{\mu\nu}u^\mu k^\nu$ where $u^\mu\equiv u^t k^\mu$, $u^t=dt/d\tau$ and $k^\mu\equiv \partial_t +dr/dt\partial_r +d\phi/dt \partial_\phi$. [Note that in the present eccentric case the so-defined $k^\mu=u^\mu/u^t$ is no longer a Killing vector.] As already mentioned,  we consider, in Eq. \eqref{delta_U1}, $\delta U$ as a function of the inverse dimensionless semi-latus rectum $u_p \equiv 1/p$ and eccentricity  $e$ (in lieu of $m_2\Omega_r$,  $m_2\Omega_\phi$) of the {\it unperturbed} orbit, as is allowed in a first-order self-force quantity. In addition, $U_0$ denotes the proper-time average of $u^t=dt/d\tau$ along the unperturbed orbit, i.e., the ratio $U_0={T_r}/{{\mathcal T}_r}|_{\rm unperturbed}$. It is approximately given by Eq. \eqref{U0} above.

For the present computation we follow the standard Teukolsky perturbation scheme as discussed in detail in Ref. \cite{vandeMeent:2015lxa}.
The expansion of the Teukolsky source-terms (which originally contain $\delta(r-r_0(t))$ and at most two of its derivatives) in powers of $e$ generates, at order $e^2$, up to four derivatives of $\delta(r-m_2/p)$ in the even part and up to three in the odd part.
This expansion gives rise to multiperiodic coefficients in the source terms, involving the combined frequencies
\beq
\omega_{m,n}=m\Omega_{\phi 0}+n \Omega_{r 0}
\eeq
with $n=0,\pm 1, \pm 2$ when working as we do up to order $e^2$.

Our computed quantity $\langle h_{uk}\rangle_t$ is regularized by subtracting its PN-analytically computed large-$l$ limit $B$, whose expansion is given by

\begin{eqnarray}
B(u_p, e, \hat a) &=&\sum_{i,j=0}^\infty e^i {\hat a}^j B^{(e^i,a^j)}(u_p)\nonumber\\
&=& B^{(e^0,a^0)}+e^2 B^{(e^2,a^0)}\nonumber\\
&+& \hat a B^{(e^0,a^1)}+\hat a^2B^{(e^0,a^2)}\nonumber\\
&+& e^2 \hat aB^{(e^2,a^1)}+e^2 \hat a^2B^{(e^2,a^2)}+\ldots\,,
\end{eqnarray}
with

\begin{widetext}
\begin{eqnarray}
-B^{(e^2,a^1)}&=&  
2u_p^{5/2}+22u_p^{7/2}+\frac{4945}{32}u_p^{9/2}+\frac{35747}{32}u_p^{11/2}+\frac{65494129}{8192}u_p^{13/2}+\frac{459731033}{8192}u_p^{15/2}+\frac{202677538545}{524288}u_p^{17/2}\nonumber\\
&&
+\frac{1374029163573}{524288}u_p^{19/2}\,,\nonumber\\
-B^{(e^2,a^2)}&=&
-\frac{7}{4}u_p^3-\frac{201}{8}u_p^4-\frac{76689}{256}u_p^5-\frac{3082555}{1024}u_p^6-\frac{1783431907}{65536}u_p^7-\frac{30088782497}{131072}u_p^8-\frac{7730696811901}{4194304}u_p^9\,.
\end{eqnarray}
\end{widetext}

As usual the low multipoles ($l=0,1$) have been computed separately, as in Eq. (138) of Ref. \cite{vandeMeent:2015lxa}. The corresponding (already subtracted) contributions to $\delta U$ are the following
\begin{widetext}
\begin{eqnarray}
-\delta U^{(e^2,a^1)}_{l=0,1}&=&
\frac{1}{2}u_p^{5/2}-7u_p^{7/2}-\frac{4865}{32}u_p^{9/2}-\frac{91135}{64}u_p^{11/2}-\frac{90124849}{8192}u_p^{13/2}-\frac{1289612077}{16384}u_p^{15/2}-\frac{284323625361}{524288}u_p^{17/2}\nonumber\\
&&
-\frac{3835930846503}{1048576}u_p^{19/2}
\,,\nonumber\\
-\delta U^{(e^2,a^2)}_{l=0,1}&=&
-\frac{1}{4}u_p^3+16u_p^4+\frac{71977}{256}u_p^5+\frac{3451661}{1024}u_p^6+\frac{2167972867}{65536}u_p^7+\frac{4765548587}{16384}u_p^8+\frac{10013809794069}{4194304}u_p^9\,.
\end{eqnarray}
\end{widetext}

We have analytically computed $\delta U(u_p,e,\hat a)$ at second order in both eccentricity $e$ and spin parameter $\hat a$ and up to order $O(u_p^{19/2})$, which corresponds to the 8.5PN order
in $\delta U$. [The {\it fractional} PN accuracy of our results for  $\delta U^{(e^2,a^1)}$ and $\delta U^{(e^2,a^2)}$ is lower because the leading-order
terms in these contributions are of order $O(u_p^{5/2})$ and $O(u_p^{3})$, respectively.]
Like in our previous works \cite{Bini:2015bfb,Bini:2016qtx} (but with the replacement of Regge-Wheeler-Zerilli perturbation theory by Teukolsky perturbation theory as 
in Ref. \cite{vandeMeent:2015lxa}, see Appendix) we combine a standard PN expansion scheme for high values of the multipole degree $l$ with the
Mano-Suzuki-Takasugi \cite{Mano:1996mf,Mano:1996vt} hypergeometric-expansion technique for lower values of $l$ (here it was used through the multipole order $l=5$).

Our new results for the Detweiler-Barack-Sago gauge invariant redshift function along eccentric orbits in a Kerr spacetime are contained in the following two contributions to the eccentricity-spin decomposition \eqref{dUea} of $\delta U(u_p, e, \hat a)$:
\begin{widetext} 
\begin{eqnarray} \label{dUe2a}
-\delta U^{(e^2,a^1)}&=&  \frac72 u_p^{5/2}+4 u_p^{7/2}-\frac{287}{2} u_p^{9/2}+\left(-\frac{5876}{3}+\frac{569}{64}\pi^2\right)u_p^{11/2}\nonumber\\
&&+\left(-\frac{1237333}{75}+\frac{122071}{512}\pi^2-\frac{4832}{15}\ln(u_p)+1728\ln(2)-2916\ln(3)-\frac{9664}{15}\gamma  \right) u_p^{13/2}\nonumber\\
&&+\left(-\frac{10843142833}{44100}+\frac{86233969}{6144}\pi^2-\frac{21874}{35}\ln(u_p)-\frac{2430}{7}\ln(3)-\frac{932332}{105}\ln(2)-\frac{8212}{7}\gamma  \right) u_p^{15/2}\nonumber\\
&&-\frac{1010822}{525}\pi u_p^8\nonumber\\
&&+\left(\frac{29277772}{2835}\gamma-\frac{5553279}{140}\ln(3)+\frac{61294612}{405}\ln(2)+\frac{341496264211}{1769472}\pi^2-\frac{547984649}{262144}\pi^4  \right. \nonumber\\
&& \left. -\frac{9765625}{324}\ln(5)+\frac{13441382}{2835}\ln(u_p)-\frac{505970041387}{198450}  \right) u_p^{17/2}\nonumber\\
&& +\frac{39743066}{33075}\pi u_p^9\nonumber\\
&&+\left(-\frac{115503655324}{363825}\gamma-\frac{311622308433}{308000}\ln(3)+\frac{1073412430012}{5457375}\ln(2)-\frac{134912}{3}\zeta(3)\right. \nonumber\\
&&+\frac{6663086579499389}{4954521600}\pi^2-\frac{1736116121221}{125829120}\pi^4+\frac{3020976}{25}\gamma\ln(3)+\frac{3020976}{25}\ln(3)\ln(2)\nonumber\\
&&-\frac{620608}{175}\ln(2)\gamma+\frac{9278816}{315}\gamma^2-\frac{101166368}{1575}\ln(2)^2+\frac{1510488}{25}\ln(3)^2+\frac{13345703125}{66528}\ln(5)\nonumber\\
&&-\frac{59233073582}{363825}\ln(u_p)-\frac{310304}{175}\ln(2)\ln(u_p)+\frac{9278816}{315}\ln(u_p)\gamma+\frac{1510488}{25}\ln(u_p)\ln(3)\nonumber\\
&& \left.+\frac{2319704}{315}\ln(u_p)^2-\frac{1311110962921933363}{75639217500}\right) u_p^{19/2}+O_{\ln{}}(u_p^{10})
\end{eqnarray}
and
\begin{eqnarray} \label{dUe2a2}
-\delta U^{(e^2,a^2)}&=&  -u_p^3+\frac{31}{2} u_p^4+356 u_p^5+\left(\frac{14378}{3}-\frac{4403}{1024}\pi^2 \right) u_p^6\nonumber\\
&& +\left(\frac{1254047}{25}-\frac{164669}{1024}\pi^2+208\gamma+104\ln(u_p)-\frac{2416}{5}\ln(2)+\frac{4374}{5}\ln(3)  \right) u_p^7\nonumber\\
&& +\left(\frac{54093631}{175}+\frac{2363953949}{196608}\pi^2+\frac{574228}{105}\gamma-\frac{389924}{35}\ln(2)+\frac{287114}{105}\ln(u_p)+\frac{222831}{10}\ln(3)\right)u_p^8\nonumber\\
&&+\frac{67303}{175}\pi u_p^{17/2}\nonumber\\
&&+\left(\frac{105145912}{2835}\gamma+\frac{36415737}{280}\ln(3)-\frac{156704768}{2835}\ln(2)+\frac{10464}{5}\zeta(3)+\frac{79088667924941}{154828800}\pi^2\right. \nonumber\\
&& \left. +\frac{7126992803}{33554432}\pi^4+\frac{48828125}{4536}\ln(5)+\frac{59348228}{2835}\ln(u_p)-\frac{598237152827}{396900}  \right) u_p^9\nonumber\\
&& +\frac{328245443}{22050}\pi u_p^{19/2}+O_{\ln{}}(u_p^{10})\,.
\end{eqnarray}
\end{widetext}

\section{Discussion}

We have improved the knowledge of the Detweiler-Barack-Sago redshift invariant (for an eccentric orbit around a Kerr spacetime) 
by providing the first analytic computation of contributions mixing eccentricity and spin effects. More precisely, in terms of
the expansion Eq. \eqref{dUea} of the first-self-force-order (inverse, average) redshift $\delta U$ in powers of eccentricity $e$
and spin parameter $\hat a$, we have computed the PN-expansions of the contributions $e^2 \hat a\delta U^{(e^2,a^1)}(u_p)$ and 
$e^2 {\hat a}^2\delta U^{(e^2,a^2)}(u_p)$ up to order $ O(u_p^{9.5})$ included, see Eqs. \eqref{dUe2a}, \eqref{dUe2a2}.

At this stage, we cannot meaningfully compare these analytical results to numerical self-force computations, because the only
extant numerical self-force computations for eccentric motions around a Kerr black hole are the sparse data listed in Table V of 
a recent work  by M. van de Meent and A. Shah \cite{vandeMeent:2015lxa}. Those numerical
data concern only very high spin parameters $\hat a= \pm 0.9$, medium-size eccentricities $e=0.1, 0.2, 0.3, 0.4$,
and, most unfortunately, are non-horizontally sampled in $p$: namely, there are no data
corresponding to the same values of  $p$ (or $u_p$) but different values of $\hat a$ and $e$. One cannot therefore
appproximately extract from these data quantities directly related to our analytical results. [The situation was
different in the case of the spin-dependence, for zero eccentricity, where we could (in Ref. \cite{Bini:2015xua})
extract dynamically useful spin-dependent information from numerical self-force data (in Ref. \cite{Shah:2012gu}) 
on $\delta U(p, e=0, \hat a)$ computed for a few values of
the spin, but (partially) horizontally sampled values of  $p$.
We think, however, that our analytical results might be useful both for checking existing Kerr self-force codes,
and for allowing the extraction of further, uncomputed PN coefficients. This is why we decided to publish them.

In future work, we intend to complete our analytical work by transcribing our results within the effective one-body
formalism \cite{Buonanno:1998gg,Buonanno:2000ef,Damour:2000we,Damour:2001tu}, by using 
the first law of binary mechanics \cite{LeTiec:2011ab,Blanchet:2012at,Tiec:2015cxa}. This will allow us to
confer a direct dynamical significance to our results Eqs. \eqref{dUe2a}, \eqref{dUe2a2}.

\subsection*{Acknowledgments}
D.B. thanks the Italian INFN (Naples) for partial support and IHES for hospitality during the development of this project.
All  the authors are grateful to ICRANet for partial support.

\appendix

\begin{widetext}

\section{A short review of the computation of the metric perturbation (from \cite{Shah:2012gu})}

Let us consider the Kerr spacetime metric (\ref{kerrmet}) with  signature switched from $+2$ to $-2$, in order to apply the standard tools of the Newman-Penrose (NP) formalism.
A principal NP frame (also termed Kinnersley frame) is the following
\begin{eqnarray}
l&=&\frac{1}{\Delta}[(r^2+a^2)\partial_t + \Delta \partial_r + a\partial_\phi]\nonumber\\
n&=&\frac{1}{2\Sigma}[(r^2+a^2)\partial_t - \Delta \partial_r + a\partial_\phi]\nonumber\\
m&=& -\frac{\bar \rho}{\sqrt{2}}\left[ia\sin\theta \partial_t +\partial_\theta +\frac{i}{\sin\theta}\partial_\phi \right]\,,
\end{eqnarray}
with nonvanishing spin coefficients 
\[
\begin{array}{cccccccccccccc}
&\rho  &=& -\displaystyle\frac{1}{r-ia\cos\theta} \qquad
&\beta &=&-\displaystyle\frac{\bar \rho \cos\theta}{2\sqrt{2}\sin\theta} \qquad 
&\pi   &=& \displaystyle\frac{ia\sin\theta\rho^2}{\sqrt{2}}\qquad 
&\tau  &=& -\displaystyle\frac{ia\sin\theta}{\sqrt{2}\Sigma} \cr
&&&&&&&&&&&\cr
&\mu  &=& \displaystyle\frac{\Delta \rho}{2\Sigma} \qquad
&\gamma &=& \mu +\displaystyle\frac{r-M}{2\Sigma} \qquad
&\alpha &=& \pi -\bar \beta\,.
&& 
\end{array}
\]
An alternative notation for the frame vectors is  $e_1=l$, $e_2=n$, $e_3=m$ and $e_4=\bar m$.  The associated frame derivatives are also denoted
\beq
D=l^\mu\partial_\mu\,,\qquad \Delta =n^\mu \partial_\mu\,,\qquad \delta =m^\mu \partial_\mu\,.
\eeq

The Teukolsky equation for a field of spin-weight $s$ in its complete form is written (symbolically) as
\beq
{\mathcal T}_s (\psi_s) =4\pi \Sigma T_s
\eeq
with
\begin{eqnarray}
{\mathcal T}_s &=& \left[\frac{(r^2+a^2)^2}{\Delta}-a^2\sin^2\theta\right]\partial_{tt}-2s\left[ \frac{M(r^2-a^2)}{\Delta}-r-ia\cos\theta \right]\partial_t +\frac{4aMr}{\Delta}\partial_{t\phi}-\Delta^{-s}\partial_r (\Delta^{s+1}\partial_r)\nonumber\\
&& -\frac{1}{\sin \theta}\partial_\theta (\sin \theta \partial_\theta)-2s\left[\frac{a(r-M)}{\Delta}+i\frac{\cos\theta}{\sin^2\theta}  \right]\partial_\phi+\left(\frac{a^2}{\Delta}-\frac{1}{\sin^2\theta}\right)\partial_{\phi\phi}+(s^2\cot^2\theta -s)\,.
\end{eqnarray}
Separation of variables 
\beq
\label{sep}
\psi_s= \sum_{lm\omega}{}_s R_{lm\omega}(r)\,\, {}_sS_{l m \omega}(\theta) \,\, e^{i(m\phi-\omega t)}\,,
\eeq
leads to the following angular (homogeneous) and radial (inhomogeneous) equations 
\begin{eqnarray}
\left\{\frac{1}{\sin\theta}\frac{d}{d\theta} \left( \sin\theta \frac{d}{d\theta} \right)+\left[\xi^2 \cos^2\theta -2s\xi \cos\theta-\frac{2ms\cos\theta+s^2+m^2}{\sin^2\theta}+E_{(l,m,s;\xi)}\right]\right\}{}_sS_{lm\omega}(\theta)&=&0
\,,\nonumber\\
\mathscr{L}_r{}_sR_{lm\omega }(r)\equiv
\left\{\Delta^{-s}\frac{d}{dr} \left(\Delta^{s+1} \frac{d}{dr} \right) +\left[\frac{K^2-2is (r-M)K}{\Delta}+4is\omega r -\lambda\right]\right\}{}_sR_{lm\omega }(r)&=&-8\pi T_{s l m \omega}\,,
\end{eqnarray}
where $\xi=a\omega$, $K=(r^2+a^2)\omega -m a$ and $\lambda\equiv \lambda_{lms;\xi}=E_{lms;\xi}-s(s+1)-2m\xi +\xi^2$, with 
\beq
E_{(l,m,s;\xi)}=l(l+1)-\frac{2s^2m }{l(l+1)}\xi +[H(l+1)-H(l)-1]\xi^2 +O(\xi^3)\,,
\eeq
being
\beq
\label{H_def}
H(l)=\frac{2(l^2-m^2)(l^2-s^2)^2}{(2l-1)l^3 (2l+1)}\,,\qquad l\ge 2\,.
\eeq 

The Teukolsky radial equation has source terms which depend on the spin-weight parameter.
In the case $s=+2$ (i.e., for $\psi_{s=2}=\psi_0$), we have in general
\beq
T_{s=2}= {\mathcal L}_1\left({\mathcal L}_2 (T_{13})-{\mathcal L}_3 (T_{11})\right)
+ {\mathcal L}_4\left({\mathcal L}_5(T_{13})-{\mathcal L}_6(T_{33})\right)\,,
\eeq
where
\[
\begin{array}{lll}
{\mathcal L}_1= \delta +\bar \pi -\bar \alpha -3\beta -4 \tau &\qquad 
{\mathcal L}_2 =D-2\epsilon -2\bar \rho\nonumber &\qquad
{\mathcal L}_3=\delta +\bar \pi -2\bar \alpha -2\beta\cr
{\mathcal L}_4= D-3\epsilon +\bar \epsilon -4\rho -\bar \rho &\qquad
{\mathcal L}_5=\delta +2 \bar \pi -2\beta &\qquad
{\mathcal L}_6=D-2\epsilon +2\bar \epsilon -\bar \rho\,,
\end{array}
\]
and $T_{11}=T_{ll}$, $T_{13}=T_{lm}$, $T_{33}=T_{mm}$ are the frame components of the stress-energy tensor of the particle with 4-velocity $u^\mu=dx^\mu/d\tau$, i.e., 
\beq
T^{\mu\nu}=\frac{\mu}{u^tr^2}u^\mu u^\nu\delta_3\,,\qquad
\delta_3=\delta(r-r_0(t))\delta(\theta-\pi/2)\delta(\phi-\phi_0(t))\,,
\eeq
given by
\begin{eqnarray}
T_{11}&=&\frac{\mu}{u^tr^2}\left(u^t-au^\phi-\frac{r^2}{\Delta}u^r\right)^2\delta_3\,,\nonumber\\
T_{13}&=&\frac{i\mu}{\sqrt{2}u^tr^3}\left(u^t-au^\phi-\frac{r^2}{\Delta}u^r\right)[au^t-(r^2+a^2)u^\phi]\delta_3\,,\nonumber\\
T_{33}&=&-\frac{\mu}{2u^tr^4}[au^t-(r^2+a^2)u^\phi]^2\delta_3\,.
\end{eqnarray}
Following the notation of Ref. \cite{Shah:2012gu}, we can write
\beq
T_{s=2}\equiv T^{(0)}+T^{(1)}+T^{(2)}\,,
\eeq
where
\beq
T^{(0)}=-{\mathcal L}_1{\mathcal L}_3 T_{11}\,,\qquad
T^{(1)}=({\mathcal L}_1{\mathcal L}_2+{\mathcal L}_4{\mathcal L}_5) T_{13}\,,\qquad
T^{(2)}=-{\mathcal L}_4{\mathcal L}_6 T_{33}\,.
\eeq

\subsection{Green's function}

One computes the Green's function of the radial equation, $G_{lm}(r,r')$ solution of the
equation
\beq
\mathscr{L}_r(G_{lm}(r,r'))=\frac{1}{\Delta}\delta(r-r')\,,
\eeq
which has the form
\begin{eqnarray}
G_{lm}(r,r')&=&\frac{(\Delta')^2}{W_{lm}}   \left[R_{\rm in}(r)R_{\rm up}(r')H(r'-r)+R_{\rm in}(r')R_{\rm up}(r)H(r-r')  \right]\nonumber\\
&\equiv& \frac{(\Delta')^2}{W_{lm}}   R_{\rm in}(r_<)R_{\rm up}(r_>) \,,
\end{eqnarray}
where $R_{\rm in}(r)$ and $R_{\rm up}(r)$ are two independent solutions to the homogeneous radial Teukolsky equation having the correct behavior at the horizon and at infinity, respectively, and $W_{lm}$ is the associated (constant) Wronskian.
The full Green's function then turns out to be
\beq
G(x,x')=\sum_{l,m}\frac{[\Delta']^2}{W_{lm}}R_{\rm in}(r_<)R_{\rm up}(r_>)\, {}_2S_{lm}(\theta){}_2  S_{lm}(\theta')e^{im(\phi-\phi')}\,.
\eeq

\subsection{Source terms}

By using the full Green's function one can solve the Teukolsky equation for $\psi_0$ ($s=2$)
\begin{eqnarray}
\psi_0 &=&-8\pi \int \Sigma' T(x',x_0)G(x,x')dr' d (\cos \theta')d\phi'\nonumber\\
&=&-8\pi \int \Sigma' [T^{(0)}+T^{(1)}+T^{(2)}]G(x,x')dr' d (\cos \theta')d\phi'\nonumber\\
&\equiv&\psi_0^{(0)}+\psi_0^{(1)}+\psi_0^{(2)}\,.
\end{eqnarray}
The coefficients $\psi_0^{(0,1,2)}$ can be computed straightforwardly and for each of them one has a left part $\psi_0^{(0,1,2)-}$ and a right one $\psi_0^{(0,1,2)+}$, i.e.,
\beq
\psi_0^{(0,1,2)}=\sum_{lm} [\psi_{0,lm}^{(0,1,2)-}H(r_0-r)+\psi_{0,lm}^{(0,1,2)+}H(r-r_0)]\, {}_2S_{l m}(\theta) e^{i(m\phi-\omega t)}\,. 
\eeq
The harmonic decomposition of $\psi_0^\pm$ is then 
\beq
\label{psi0harmdec}
\psi_0^\pm= \sum_{lm}{}_2 R_{lm\omega}^\pm(r)\,\, {}_2S_{l m \omega}(\theta) \,\, e^{i(m\phi-\omega t)}\,,
\eeq
with 
\beq
{}_2 R_{lm\omega}^\pm(r)=\psi_{0,lm}^{(0)\pm}+\psi_{0,lm}^{(1)\pm}+\psi_{0,lm}^{(2)\pm}\,,
\eeq
leading to
\beq
\label{Rpmdef}
{}_2 R_{lm\omega}^-(r) ={\mathscr A}_{lm\omega, \rm (up)}^-(r_0) R_{\rm in}(r)\,,\qquad
{}_2 R_{lm\omega}^+(r) ={\mathscr A}_{lm\omega, \rm (in)}^+(r_0) R_{\rm up}(r)\,.
\eeq
The coefficients ${\mathscr A}_{lm\omega, \rm (up)}^-$ and ${\mathscr A}_{lm\omega, \rm (in)}^+$ can be expressed (formally) as 
\beq
\label{quasi_finale}
 {\mathscr A}_{lm\omega, \rm (up)}^-=\frac{1}{W_{lm}}\left[\alpha_{lm}^- R'_{\rm up}(r_0) + \beta_{lm}^- R_{\rm up}(r_0)\right]\,,\qquad
 {\mathscr A}_{lm\omega, \rm (in)}^+=\frac{1}{W_{lm}}\left[\alpha_{lm}^+ R'_{\rm in}(r_0) + \beta_{lm}^+ R_{\rm in}(r_0)\right]\,.
\eeq

\subsection{Hertz potential}

To compute the perturbed metric one introduces the Hertz-Debye potential $\Psi$, which is related to $\psi_0$ by \cite{Shah:2012gu}
\beq
\label{Psidef}
\psi_0=\frac18 \left[  {\mathcal L}^4 \bar \Psi+12 M \partial_t \Psi  \right]\,,
\eeq
with
\beq
{\mathcal L}^4={\mathcal L}_1{\mathcal L}_0{\mathcal L}_{-1}{\mathcal L}_{-2}\,,\qquad {\mathcal L}_s=-[\partial_\theta -s\cot\theta +i\csc  \theta \partial_\phi]-ia\sin\theta \partial_t
\,.
\eeq
The harmonic decompositions of $\Psi$ and its complex conjugate $\bar\Psi$ are given by
\beq
\Psi 
= \sum_{lm\omega}{}_2 {\mathcal R}_{lm\omega}(r)\,\, {}_2S_{l m \omega}(\theta) \,\, e^{i(m\phi-\omega t)}\,,\qquad
\bar \Psi 
= \sum_{lm\omega} (-1)^m \, {}_2 \bar{{\mathcal R}}_{l,-m,-\omega}(r)  \,\, {}_{-2}S_{lm\omega}(\theta)\,\, e^{i(m\phi-\omega t)}\,,
\eeq
respectively.
The Teukolsky-Starobinski identity
\beq
{\mathcal L}^4 \left( {}_{-2}S_{lm\omega}e^{i(m\phi-\omega t)}  \right)=D \,\, \,\left( {}_{2}S_{lm\omega}e^{i(m\phi-\omega t)}\right)\,,
\eeq
with 
\beq
D^2=\lambda_{\rm CH}^2 (\lambda_{\rm CH}+2)^2+8 a \omega \lambda_{\rm CH}(m-a\omega)(5\lambda_{\rm CH}+6)+48a^2\omega^2 [2\lambda_{\rm CH} +3(m-a\omega)^2]\,,
\eeq
and
$\lambda_{\rm CH}=E_{(l,m,2;\xi)}+\xi^2 -2m\xi -2$ is the Chandrasekhar constant,
implies
\beq
{\mathcal L}^4  (\bar \Psi ) = \sum_{l,m} (-1)^m \, {}_2 \bar{{\mathcal R}}_{l,-m,-\omega}(r)  \,\, D\,\, {}_{2}S_{lm\omega}   e^{i(m\phi-\omega t)}\,.
\eeq
Up to the second order in $a$ we have
\begin{eqnarray}
D &=& l(l-1)(l+2)(l+1) -4 (l-1) (l+2)  m\omega a\nonumber\\
&+&\frac{4(l-1)(l+2)}{(2l+3)(2l-1)l^2(l+1)^2}(l^6+3l^5+5l^4m^2-9l^4-23l^3+10m^2l^3-12l^2+19m^2l^2+14m^2l+12m^2)\omega^2a^2\nonumber\\
&+& O(a^3)\,.
\end{eqnarray}
Taking into account that $\partial_t \Psi = -i\omega \Psi$, Eq. (\ref{Psidef}) thus becomes
\beq
\psi_0 =\sum_{l,m} \frac18 \left[ (-1)^m  D\,\,  {}_2 \bar{{\mathcal R}}_{l,-m,-\omega}(r)  -12 iM\omega {\mathcal R}_{lm\omega}(r) \right]{}_{2}S_{lm\omega}   e^{i(m\phi-\omega t)}\,.
\eeq
Recalling then the harmonic decomposition (\ref{psi0harmdec}) of $\psi_0$ implies
\beq
R_{l,m,\omega}=\frac18 \left[ (-1)^m  D\,\,  {}_2 \bar{{\mathcal R}}_{l,-m,-\omega}(r)  -12 iM\omega {\mathcal R}_{lm\omega}(r) \right]\,,
\eeq
which once inverted yields
\beq
{}_2{\mathcal R}{}_{lm\omega}=8 \frac{(-1)^m D}{D^2+144M^2\omega^2}  {}_2\bar R_{l,-m,-\omega} \,\, + 8 \frac{12i M\omega }{D^2+144M^2\omega^2}{}_2R_{l,m,\omega} \,,
\eeq
what is needed to compute $\Psi$.

\subsection{Metric reconstruction}

The radiative ($l\ge 2$) perturbed metric (up to parts for which $\psi_0$ vanishes) is given by
\begin{eqnarray}
h_{\alpha\beta}=\rho^{-4}\{
n_\alpha n_\beta D_{nn} +\bar m_\alpha \bar m_\beta D_{\bar m\bar m} - n_{(\alpha} \bar m_{\beta )} D_{n \bar m}\}\Psi +{\rm c.c.}\,,
\end{eqnarray}
where
\begin{eqnarray}
D_{nn}&=& (\bar \delta -3\alpha -\bar \beta +5 \pi)(\bar \delta-4\alpha +\pi)\nonumber\\
D_{\bar m\bar m}&=& (\Delta +5 \mu -3\gamma +\bar \gamma)(\Delta +\mu -4\gamma)\nonumber\\
D_{n \bar m}&=& (\bar \delta -3\alpha +\bar \beta +5 \pi+\bar \tau)(\Delta +\mu -4\gamma)+(\Delta+5\mu -\bar \mu -3\gamma -\bar \gamma)(\bar \delta -4\alpha +\pi)\,. 
\end{eqnarray}

On the other hand, the contribution of the non-radiative modes $l=0,1$ comes from the change in mass and angular momentum due to the presence of the orbiting particle of mass $\mu$.
The Kerr metric perturbed in mass and angular momentum (in BL coordinates) acquires the following nonzero components (for $r>r_0$)
\beq
h_{tt}=-\frac{2\delta M}{r}\,,\quad 
h_{rr}=-\frac{2r^2}{M\Delta^2}[(Mr+a^2)\delta M-a\delta J]\,,\quad 
h_{\phi\phi}=\frac{2a}{Mr}[(r+M)a\delta M-(r+2M)\delta J]\,,\quad 
h_{t\phi}=\frac{2\delta J}{r}\,,
\eeq
with $\delta M=E=\mu u_t$ and $\delta J=L=-\mu u_\phi$.

Finally, one computes the gauge-invariant Detweiler-Sago redshift variable \eqref{delta_U1} with 
\begin{eqnarray}
h_{uk}&=&\frac{1}{\rho^{4}u^t}\{
(n\cdot u)^2 D_{nn} +(\bar m \cdot u)^2  D_{\bar m\bar m} - (n\cdot u)(\bar m \cdot u) D_{n \bar m}\}\Psi +{\rm c.c.}\,.
\end{eqnarray}

\end{widetext}

\end{document}